\documentclass[aps, prd, superscriptaddress, nofootinbib, twocolumn]{revtex4-2}

\usepackage{amsmath}
\usepackage{graphicx}
\usepackage{dcolumn}
\usepackage{bm}
\usepackage{wasysym}
\usepackage{hyperref}
\usepackage[mathlines]{lineno}
\usepackage{xcolor}

\usepackage{gensymb}
\usepackage{enumitem}
\usepackage{multirow}
\usepackage{booktabs}
\usepackage{array}
\newcolumntype{P}[1]{>{\centering\arraybackslash}p{#1}}
\usepackage{amssymb}
\usepackage{adjustbox}
\usepackage{orcidlink}
\usepackage{pifont}

\usepackage[normalem]{ulem}

\usepackage{xcolor, soul}

\newcommand{\du}{\mathrm{d}}   
\newcommand*{\defeq}{\mathrel{\vcenter{\baselineskip0.5ex \lineskiplimit0pt
                     \hbox{\scriptsize.}\hbox{\scriptsize.}}}%
                     =}

\begin{document}

\title{Assessing the stability of ultracompact spinning boson stars with nonlinear evolutions}


\author{Tamara Evstafyeva
\orcidlink{0000-0002-2818-701X}}
\email{tevstafyeva@perimeterinstitute.ca}
\affiliation{Perimeter Institute for Theoretical Physics,
Waterloo, Ontario N2L 2Y5, Canada}
\author{Nils Siemonsen
\orcidlink{0000-0001-5664-3521}}
\email{nils.siemonsen@princeton.edu}
\affiliation{Princeton Gravity Initiative, Princeton University, Princeton, NJ 08544, USA}
\affiliation{Department of Physics, Princeton University, Princeton, NJ 08544, USA}
\author{William E. East
\orcidlink{0000-0002-9017-6215}}
\affiliation{Perimeter Institute for Theoretical Physics,
Waterloo, Ontario N2L 2Y5, Canada}
\date{\today}

\begin{abstract}

We reinvestigate the stability properties of ultracompact spinning boson stars with a stable light ring using fully nonlinear 3+1 and 2+1 numerical relativity simulations and two different formulations of the Einstein equations. We find no evidence of an instability on timescales of $t \mu
\sim 10^4$ (in units of the scalar mass), when allowing the star to be perturbed either solely by discretization error or by imposing various types of perturbations to our initial data. 
We find that the initially imposed perturbations exhibit slow decay, even for magnitudes just below the order where immediate collapse is induced.
\end{abstract}

\maketitle

\section{Introduction} 

The recent advancements in gravitational-wave detections by LIGO-Virgo-KAGRA~\cite{AdvancedDetectorsLVK2018,Abbott:2016blz,KAGRA:2021vkt} and electromagnetic observations, such as those of the GRAVITY collaboration \cite{GRAVITY:2018ofz} and the Event Horizon Telescope~\cite{EventHorizonTelescope:2022xqj} present increasingly strong evidence that our universe is populated by extremely compact, massive objects---interpreted within the black hole (BH) paradigm. However, both bottom up constructions and top down high-energy motivated horizonless ultracompact objects (UCOs), or \textit{BH mimickers}, have been put forward to challenge this paradigm~\cite{Cardoso:2019rvt,Bambi:2025wjx,Carballo-Rubio:2025fnc}. Generically, UCOs
exhibit (at least) two light rings (LRs)~\cite{Cunha:2017qtt}: an outer LR, where null geodesics are unstably trapped (in direct analogy to their counterparts in BH spacetimes), and an inner LR, where null geodesics are instead \textit{stably} trapped. 
In the eikonal limit, the dynamics of massless modes in the inner LR approach that of trapped null geodesics; hence, stable (unlike unstable) null trapping signals the appearance of long-lived massless modes
\cite{Keir:2014oka,Cardoso:2014sna} (also see Refs.~\cite{Benomio:2018ivy,Holzegel:2013kna} for related contexts). It has been speculated that such slow decay enables weakly nonlinear effects to accumulate leading to a \textit{non-linear `light ring'} instability~\cite{Keir:2014oka,Cardoso:2014sna}.

Several works have explored the nonlinear dynamics of stable null trapping within asymptotically flat spacetimes. In Ref.~\cite{Cunha:2022gde}, such a LR instability was reported to be present in ultracompact scalar boson stars (BSs) and Proca stars. It was concluded that the fate of these objects was either BH formation or migration to a less compact model. 
Further, recent numerical work in Ref.~\cite{Siemonsen:2024snb} found no evidence of a nonlinear instability mechanism in the case of ultracompact spinning BSs as by-products of binary collisions (though, on relatively short timescales and restricting to axisymmetry), whilst Ref.~\cite{Marks:2025jpt} confirmed nonlinear stability of ultracompact spherically symmetric BSs on timescales of $t\mu \sim 1.5 \times 10^4$ (with perturbations driven by the discretization error), where $\mu$ is the scalar mass. In Ref.~\cite{Benomio:2024lev}, and further refined in Ref.~\cite{Redondo-Yuste:2025hlv}, it was observed that a nonlinear massless scalar test-field can exhibit turbulent behavior, when trapped in stable light rings. Whether gravitational perturbations also exhibit this turbulent behavior and lead to an unstable process (analogous to say the nonlinear instability of Anti-de Sitter space through a weakly turbulent mechanism \cite{Bizon:2011gg,Jalmuzna:2011qw,Buchel:2012uh}) are still open questions.

Dynamical stability of compact objects undoubtedly plays a vital role in understanding their astrophysical relevance and helps guide gravitational-wave search campaigns. For instance, in spinning UCOs, the existence of an ergoregion implies the presence of the \textit{linear} ergoregion instability, which typically operates on the timescales of $\gtrsim 10^5M$ (where $M$ is the total mass)~\cite{Friedman:1978,Vilenkin:1978uc,Comins:1978,Maggio:2018ivz,Moschidis:2016zjy,Zhong:2022jke}. While the end-state of this instability is largely unknown, it has been used to argue against the existence of several classes of highly spinning BH mimickers \cite{Fan:2017cfw,Barausse:2018vdb,Mastrogiovanni:2025ixe}. Analogously, the presence of a nonlinear instability due to the stable light ring would impact virtually \textit{all} proposed BH mimickers. However, fundamental questions about the putative instability mechanism remain open: (i) is there a nonlinear process driving a departure from stationarity? (ii) what is the nature of this mechanism? (iii) what is the associated timescale of the instability and its dependence on the amplitude of the perturbation? (iv) what is the end-state of this process? (v) how does this mechanism depend on the specific properties of the stable LR?

\begin{table*}[t!]
\centering
\caption{
Summary of numerical relativity codes and final setups employed in this work. For each code we specify (i) $d$ -- the number of dimensions used in the evolution, (ii) the formulation of the Einstein equations, (iii) the gauge choice, (iv) $\mu \du x$ -- the range of resolutions on the innermost level considered in this work, (iv) $\mu L$ -- the size of the computational domain, and (v) $l_{\rm{max}}$ -- the maximum number of refinement levels. We note that for the {\sc GHC} code $\mu L = \infty$ denotes the compactified computational domain.
} 
\begin{tabular}{| c | c | c | c | c | c | c |} 
\hline
~~~Code~~~& ~~~$d$~~~&~~~Formulation~~~&~~~Gauge~~~&~~~$\mu \du x$~~~&~~~$\mu L$~~~&~~~$l_{\rm{max}}$~~~\\ 
\hline
{\sc ExoZvezda} & 3 & CCZ4 & Moving puncture gauge & 0.08-0.25 & 1024 & 7 \\ 
\hline
{\sc GHC} & 3 & Generalized harmonic & Stationary gauge & 0.15-0.3 & $\infty$ & 7 \\
\hline 
{\sc GHC} & 2 & Generalized harmonic & Stationary gauge & 0.004-0.0015 & $\infty$ & 8
\\ 
\hline
\end{tabular}
\label{tab:codes_setup} 
\end{table*}

In this work, we study the dynamical stability of ultracompact spinning BSs exhibiting stable LRs using fully nonlinear numerical evolutions in the Baumgarte-Shapiro-Shibata-Nakamura-Oohara-Kojima (BSSNOK), the conformal and covariant Z4 (CCZ4), and the generalized harmonic formulations of the Einstein equations. Focusing on the ultracompact scalar boson star solution reported to be unstable in Ref.~\cite{Cunha:2022gde}, we find, in contrast, no evidence for instability when considering various types of initial perturbations, which typically suffer from slow decay. 
Furthermore, we discuss challenges encountered whilst employing standard numerical relativity techniques to evolve these extreme spacetimes and summarize key ingredients that facilitated their faithful evolutions. Most importantly, we highlight possible numerical artifacts and gauge instabilities, which could be confused for physical instabilities (both linear and nonlinear) and propose simple recipes to circumvent such issues. Overall, our findings suggest that the slow decay of linear modes does not lead to an eventual instability on the timescales of our simulations.  

This paper is organized as follows. In Section~\ref{sec:intro_inst}, we start with a brief discussion of how numerical evolutions can help assess stability of stationary spacetimes. Specializing to boson stars in axisymmetry, in Sections~\ref{sec:theory}-\ref{sec:stability_overview} we present their formulation and discuss the latest results concerning stability of their various classes. Next, in Section~\ref{sec:numerics}, we review the numerical setup employed in this work to evolve ultracompact rotating boson stars and the diagnostics we employ to assess their stability numerically. In Section~\ref{sec:results}, we present our stability results for different types of initial perturbations and discuss the main challenges we have encountered when evolving these extreme models. Finally, in Section~\ref{sec:conclusions} we conclude with some remarks. In the Appendices, we include some additional results, including on numerical convergence and gauge instabilities.

In the subsequent sections, unless otherwise stated, we employ $G = c = 1$ units. 

\section{Assessing stability with nonlinear evolutions}\label{sec:intro_inst}
Numerical evolutions of the full equations of motion provide a powerful way to assess the stability of stationary spacetimes in instances where perturbation theory is unavailable. The general strategy is to begin with a stationary solution and evolve it forward in time after introducing a sufficiently small perturbation, or by letting numerical truncation error serve as the perturbation. If one finds a growing departure from the stationary spacetime, the evolution of this perturbation can be used to (i) measure the timescale, (ii) distinguish the linear or nonlinear nature, and (iii) study the eventual back-reaction of the instability. In theory, one can use this method to search for \textit{any} instability, linear or nonlinear, that is consistent with the underlying symmetries of the simulation. 

However, practically speaking, there are several limitations. In order to be able to identify an instability, one must carry on a simulation for a sufficiently long time so as to be able to measure the growth of the instability. This must be at least as long as the characteristic timescale of the instability, but could be significantly longer if the instability is only weakly excited initially and thus must grow significantly before rising above other, stable components of the perturbation to the stationary solution. Furthermore, one must resolve any length or timescale associated with the instability, which could be smaller than the scales associated with the stationary solution. Finally, one must eliminate the possibility of unphysical instabilities, which includes:
\begin{enumerate}
\item Gauge instabilities, that is, growing solutions to the evolution equations that merely correspond to changes in the spacetime coordinates without affecting the physical solution.
\item Constraint violating instabilities, that is, growing perturbations, which violate the constraints of the physical evolution system.
\item  Numerical error, including the secular accumulation of truncation error during an evolution, or an instability associated with the numerical scheme or the imposition of boundary conditions.
\end{enumerate}

There are several strategies that one can use to check that an instability found this way is not unphysical:
\begin{enumerate}
\item Monitor the growth of gauge-invariant quantities that cannot change solely due to gauge effects.
\item Check that the amount of constraint violation, measured relative to amplitude of the growing perturbation, converges to zero with increasing resolution (assuming the initial data satisfies the constraint equations). It is not uncommon for a growing physical instability to lead to constraint violation that grows at the same rate. However, the amount of constraint violation at a given amplitude for the perturbation should be controlled by the numerical resolution. 
\item Demonstrate that the timescale associated with the instability converges with resolution to a finite number. For a linear, exponentially growing mode, this would be the e-folding time, while for a nonlinear instability the timescale may have a more complicated dependence on the perturbation amplitude. A canonical example of the latter is the turbulent instability of a scalar field in Anti-de Sitter, where the growth rate was found to be proportional to the amplitude squared~\cite{Bizon:2011gg}.  
\item Verify that the instability persists when (and is consistent with) varying initial seed amplitudes. 
\end{enumerate}

In the remainder of this work, we concentrate on the particular case of rotating scalar BS solutions.
We discuss some settings where gauge and constraint violating instabilities arise, and how they can be eliminated,
but find no evidence for a physical instability across all cases and the imposed symmetry assumptions considered. 

\section{Theory}\label{sec:theory}

Boson stars are described by a massive complex scalar field $\varphi$, minimally coupled to the Einstein-Hilbert term
\begin{equation} \label{eq:action}
    S = \int \du^4 x \sqrt{-g} \left[\frac{R}{16 \pi} - g^{\mu \nu} \nabla_{\mu} \bar{\varphi} \nabla_{\nu} \varphi - V(|\varphi|^2) \right],
\end{equation}
where $V(|\varphi|^2)$ denotes the scalar potential. Here, we focus on the solitonic potential given by\footnote{We note that our convention for the action is different from that of Ref.~\cite{Marks:2025jpt}. In particular, what we call $\sigma_0$ here would be
called $\sigma_0/\sqrt{2}$ in the conventions of Ref.~\cite{Marks:2025jpt}.}
\begin{equation} \label{eq:potential}
    V(|\varphi|^2) = \mu^2 |\varphi|^2 \left(1 - \frac{2|\varphi|^2}{\sigma_0^2} \right)^2,
\end{equation}
where $\mu$ is the scalar mass and $\sigma_0$ quantifies the attractive nature of self-interactions\footnote{Smaller $\sigma_0$ implies larger self-interactions.}. The variation of Eq.~\eqref{eq:action} results in the system of Einstein-Klein-Gordon (EKG) equations
\begin{align}
    G_{\mu \nu} &= 8\pi T_{\mu \nu}, \\
    \nabla_{\alpha}\nabla^{\alpha} \varphi &= \varphi \frac{\partial V(|\varphi|^2)}{\partial |\varphi|^2},
\end{align}
which we focus on solving in this work. 

To construct BS solutions, we start with the metric in Lewis-Papapetrou coordinates 
\begin{align} \label{eq:metric}
    \du s^2 = - f \du t^2 + & lf^{-1} \{g \left(\du r^2 + r^2 \du \theta^2 \right) \nonumber  \\
    &+ r^2 \mathrm{sin}^2 \theta \left(\du \phi - \Omega r^{-1} \du t \right)^2 \},
\end{align}
where $f,\ l,\ g$, and $ \Omega$ are functions of $r$ and $\theta$, and the scalar field profile 
\begin{equation} \label{eq:scalarprofile}
    \varphi = A(r, \theta) e^{i(\omega t + m \phi)},
\end{equation}
where $\omega$ denotes the BS frequency and $m$ the azimuthal number. With the use of the ansätze in Eqs.~\eqref{eq:metric}-\eqref{eq:scalarprofile}, we then solve the EKG equations using a relaxation algorithm described in Ref.~\cite{Siemonsen:2020hcg} (see also Ref.~\cite{Kleihaus:2005me}). We restrict the discussion of the main text to two ultracompact spinning BS models, which have (at least) two light rings: (i) model $\texttt{S005}$ with $m=1$, $\sigma_0=0.05$, $\omega/\mu = 0.16$, compactness $C = 0.33$, radius $\mu R = 18.5$, ADM mass $\mu M = 6.12$, and angular momentum $J/M^2=0.68$ and (ii) model $\texttt{S02}$ with $m=3$, $\sigma_0=0.2$, $\omega/\mu=0.55$, $C=0.38$, $\mu R = 8.70$, $\mu M=3.27$ and $J/M^2=1.07$. 
However, we have evolved many other additional models that we summarize in the table of Appendix~\ref{app:other_models}. 

We note that the former (\texttt{S005}) model was previously reported to suffer from LR instability in Ref.~\cite{Cunha:2022gde} on a time-scale of $t \mu \sim 5000$ (see their Fig.~2) when perturbed with truncation error. The latter model (\texttt{S02}), on the other hand, is expected to be
subject to a linear non-axisymmetric instability in the full 3+1 setting~\cite{Siemonsen:2020hcg}. However, as we will discuss below, in axisymmetry (i.e.~2+1), \texttt{S02} is linearly stable over the relevant timescales. 

\section{Comments on stability of boson stars}\label{sec:stability_overview}

\begin{figure*}[t!]
    \includegraphics[width=\linewidth]{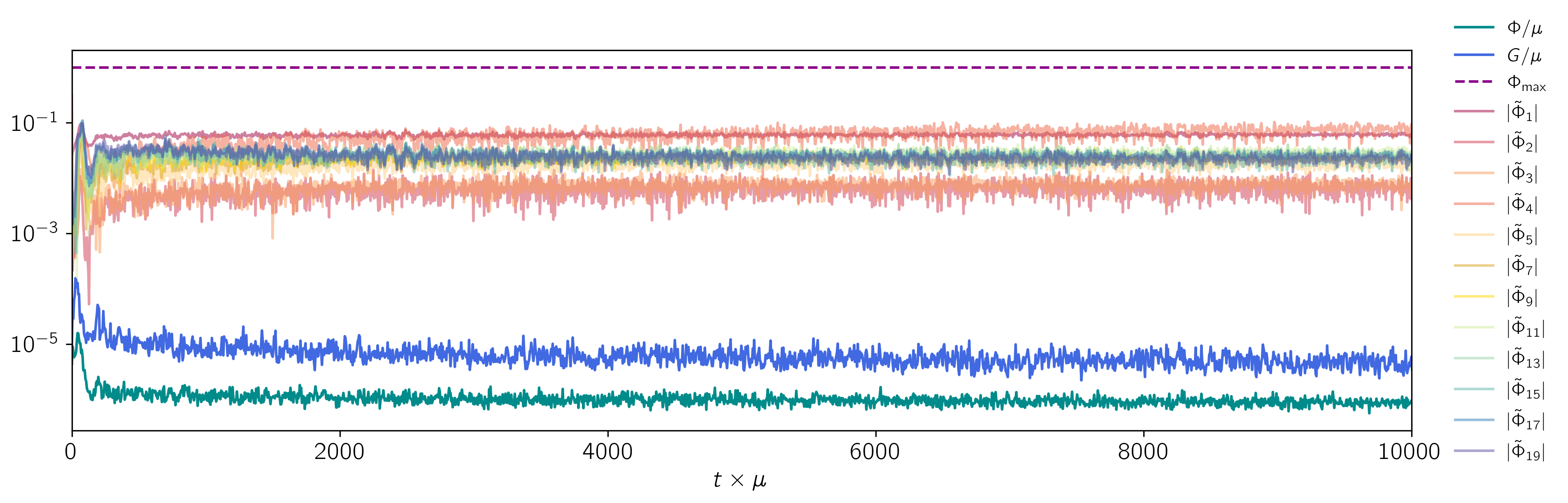}
    \caption{Illustration of the stationarity of the ultracompact spinning BS solution \texttt{S005} (perturbed only by numerical discretization error) using measures $\Phi$ and $G$, defined in Eqs.~\eqref{eq:phi_diagnostics}--\eqref{eq:gtt_diagnostics}. We also show the first few and higher-order $\tilde{m}$-modes of $\Phi$, normalized by $\Phi_0$, so that $|\tilde{\Phi}_{\tilde{m}}| = |\Phi_{\tilde{m}}|/|\Phi_0|$. The magnitude of the $\tilde{m}=4$ mode is comparable to the magnitude of the $\tilde{m}=1$, and exhibits a slight growth, which is discussed in the main text.
    }
    \label{fig:stationarity}
\end{figure*}

Stability properties of wide classes of BS models have been extensively investigated in the past, see, e.g.,~Refs.~\cite{Lee:1988av,Gleiser:1988ih,Gleiser:1988rq,Seidel:1990jh, Seidel:1990nu,Balakrishna:1997ej,Kleihaus:2011sx,Kusmartsev:1990cr, DiGiovanni:2020ror,Sanchis-Gual:2019ljs,Kain:2021rmk,Cunha:2022gde, Siemonsen:2020hcg,Alcubierre:2021mvs,Ge:2024itl,Marks:2025jpt} (see also Refs.~\cite{Liebling:2012fv,Bezares:2024btu} for a review); although the ultracompact variants of such stars have not been considered so extensively. 
In spherical symmetry, the stability was first analyzed for BSs with no self-interaction ($\sigma_0 \rightarrow \infty$)~\cite{Gleiser:1988ih,Lee:1988av,Kusmartsev:1990cr}. In particular, it has been shown that each extremum of the $M(A_{\rm{ctr}})$ curve, where $A_{\rm{ctr}} = A(0)$, corresponds to a change in the stability of a radial oscillation mode of the star (see also Fig.~4 of Ref.~\cite{Ge:2024itl} for a visual example). 

However, in the case of spinning BSs, the stability is not solely determined by locations of extrema of $M(A_{\rm ctr})$. Generally, a large portion of spinning BSs suffer from non-axisymmetric instabilities and, in particular, rotating mini BSs ($\sigma_0 \rightarrow \infty$) have been found to be unstable due to a $m=2$ dynamical bar-mode~\cite{Sanchis-Gual:2019ljs}. The stability properties of rotating stars with various potentials was later investigated in Ref.~\cite{Siemonsen:2020hcg}, finding similar instabilities driven by higher-order non-axisymmetric modes. However, the authors also showed that addition of large self-interactions to the potential [like in Eq.~\eqref{eq:potential}] can quench the instabilities in some regions of the parameter space (see~Figs.~5--6 of Ref.~\cite{Siemonsen:2020hcg}). In the non-relativistic regime, or the case of rotating Newtonian boson stars (aka Bose stars), the instability of stars with attractive or negligible self-interactions has been understood and attributed to particles transitioning to non-rotating states~\cite{Dmitriev:2021utv}. In fact, the authors of Ref.~\cite{Dmitriev:2021utv} show linear instability of these models using analytical arguments, and verify numerically that Bose stars with $m=1$ become stable for sufficiently strong repulsive self-interactions.
However, it remains unclear whether similar analytical approaches can be extended to the relativistic regime — in particular to ultracompact stars — so as to elucidate their linear instability, if present, and its underlying qualitative mechanism.

\section{Numerical setup}

\subsection{Code infrastructure}\label{sec:numerics}

We perform numerical evolutions using two independent numerical relativity codes: (i) {\sc ExoZvezda}~\cite{exozvezda}, an extension of {\sc GRChombo}~\cite{Radia:2021smk,Andrade:2021rbd}, built on the adaptive-mesh-refinement (AMR) library provided by {\sc Chombo}~\cite{chombo} and (ii) the generalized harmonic code of Refs.~\cite{Pretorius:2004jg,Siemonsen:2020hcg}, denoted by {\sc GHC} throughout.
Due to the differences in the codes, we discuss their setups, as well as numerical challenges we have encountered, separately. However, in Table~\ref{tab:codes_setup} we summarize their key ingredients. 

{\sc ExoZvezda} utilizes fourth-order finite-differencing of the CCZ4 system~\cite{Alic:2011gg}, where we fix the damping parameters to $\kappa_1 = 0.1/M$, $\kappa_2 = 0$ and $\kappa_3 = 1$ (note that our $\kappa_1$ is defined as in Eq.~(27) of Ref.~\cite{Alic:2013xsa}). We
complement the CCZ4 evolution scheme with the moving puncture gauge~\cite{Campanelli:2005dd, Baker:2005vv}, as given by Eqs.~(28)-(31) of Ref.~\cite{Radia:2021smk}. We further utilize sixth-order Kreiss-Oliger dissipation with a dissipation parameter set to 0.3 (see e.g.~Section~2.2.4. of Ref.~\cite{Clough:2015sqa}), although find our conclusions unchanged for lower values.
In all of our simulations we set up a computational domain of length $\mu L = 1024$ with 7 additional AMR levels and utilize a bitant  
symmetry along the $z$-direction (i.e. the spin axis)\footnote{Bitant symmetry implies that even (odd)  perturbations exclude odd (even) $l$-modes during the evolution.}. Our AMR is set up in such a way that the star is completely covered by the finest level. We consider resolutions in the ranges of $ \mu \du x \approx$ 0.08-0.25 on the finest level, and find that our diagnostics converge at least at third order (see~Appendix~\ref{convergence_tests} for more details). 

The {\sc GHC} code is also based on fourth-order accurate finite difference spatial discretization and Runge-Kutta time stepping, with sixth-order Kreiss-Oliger dissipation, but utilizes the generalized harmonic formulation of the Einstein equations \cite{Pretorius:2004jg}. Throughout, we stick to stationary gauge, which fixes the source functions $H^a=\Box x^a$ to be constant in time and equal to the those of the stationary boson star solution in the form \eqref{eq:metric}. Spatial infinity is included in the computational domain through a compactification of the Cartesian coordinates. Constraint damping terms, with $\kappa=2/M$ and $\rho=-0.5$, are used \cite{Gundlach:2005eh,Pretorius:2006tp}. 
For the purposes of this work, in 3+1 (2+1) evolutions we employ fixed mesh refinement with 7 (8) levels centered on the star. 
We note that in Table~\ref{tab:codes_setup} the setup of the {\sc GHC} code utilizing the generalized harmonic formulation results in gauge instabilities (we comment on this in more detail in Section~\ref{sec:gauge_generalised} and Appendix~\ref{app:gauge_instability}). The timescale of this gauge instability, however, is so long, that it is inaccessible by the 3+1 evolutions. We therefore additionally performed 2+1 evolutions with generalized harmonic formulation to investigate this further.

\subsection{Diagnostics}

During the numerical evolutions we quantify the departure from stationarity of the solution using the global maximum of the derivative of the scalar field amplitude and the $tt$-component of the spacetime metric
\begin{align} 
    \Phi &\defeq \mathrm{max} |\partial_t |\varphi|^2|, \label{eq:phi_diagnostics} \\
    G &\defeq \mathrm{max} |\partial_t g_{tt}|. \label{eq:gtt_diagnostics}
\end{align}
We also monitor the gauge-invariant quantity
\begin{equation} \label{eq:phi_max}
    \Phi_{\rm{max}} \defeq \frac{\mathrm{max}(|\varphi|)}{\mathrm{max}(|\varphi(t=0)|)},
\end{equation}
which in case of departure of the solution from its equilibrium should deviate from unity. We perform an azimuthal decomposition of $\Phi$ over the coordinate volume $\mathcal{V}$ following
\begin{equation}\label{eq:decomposition}
    \Phi_{\tilde{m}} = \int_{\mathcal{V}} |\partial_t |\varphi|^2| e^{i\tilde{m}\phi} \du^3 x,
\end{equation}
where $\Phi_{\tilde{m}}$ encodes the power of the $\tilde{m}$-th mode\footnote{We use $\tilde{m}$ here to differentiate it from $m$ that enters the scalar field ansatz in Eq.~\eqref{eq:scalarprofile}.} and we monitor the evolution of each $\Phi_{\tilde{m}}$ up to $\tilde{m} \leq 20$. Equation~\eqref{eq:decomposition} is a common diagnostic for quantifying the nature of non-axisymmetric instabilities for neutron stars and BSs~\cite{Kiuchi:2011re,East:2016zvv,Espino:2019xcl,Siemonsen:2023hko}. The discretization error (especially for low resolution runs) results in a slight linear drift across the grid of the BS at later times. To account for this drift in the computation of $\Phi_{\tilde{m}}$ modes in Eq.~\eqref{eq:decomposition}, we recompute the azimuthal angle $\phi$ with the center-of-mass coordinates of the star, which we track by fitting a Gaussian to the conformal factor. Lastly, both codes utilize slightly different diagnostics to track the Einstein constraint violations. In particular, in {\sc ExoZvezda} we track volume-weighted $L^2$-norms of the Hamiltonian and momentum constraint violations [see e.g. Eqs.~(33-34) of Ref.~\cite{Radia:2021smk} for their explicit definitions]. 
\begin{align}
\begin{aligned}
    \lVert \mathcal{H} \rVert_2 &=  \sqrt{\frac{1}{\mathcal{V}} \int_{\mathcal{V}} \mathcal{H}^2 \du^3 x}, \\
    \lVert \mathcal{M} \rVert_2 &= \sqrt{ \frac{1}{\mathcal{V}} \int_{\mathcal{V}} (\mathcal{M}_x^2 + \mathcal{M}_y^2 + \mathcal{M}_z^2)\du^3 x}.
    \label{eq:ham_mom_constraints}
\end{aligned}
\end{align}
In the {\sc GHC} setting, we quantify any constraint violations with the integral norm
\begin{align}
    \tilde{C}=\frac{1}{4}\int_{B} \du ^3x\sqrt{\gamma}\left[\sum_{a=0}^3(H^a-\Box x^a)^2\right]^{1/2},
    \label{eq:gh_cnst}
\end{align}
where $\gamma$ is the determinant of the metric of the spatial slices and $B$ is a coordinate ball with a $100M$ radius.

\section{Results}\label{sec:results}

\begin{figure*}[t!]
    \includegraphics[width=0.45\linewidth]{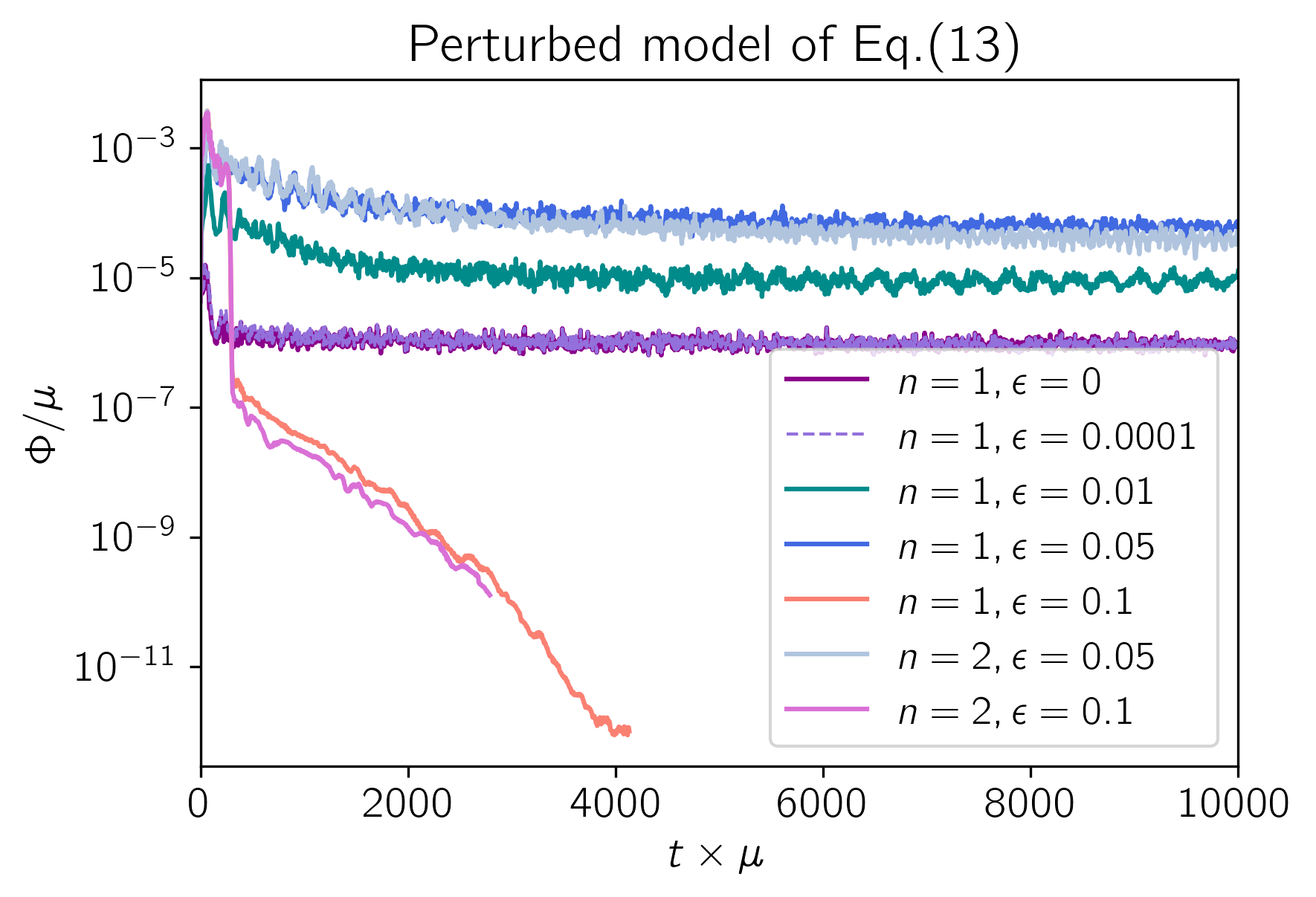}
    \includegraphics[width=0.45\linewidth]{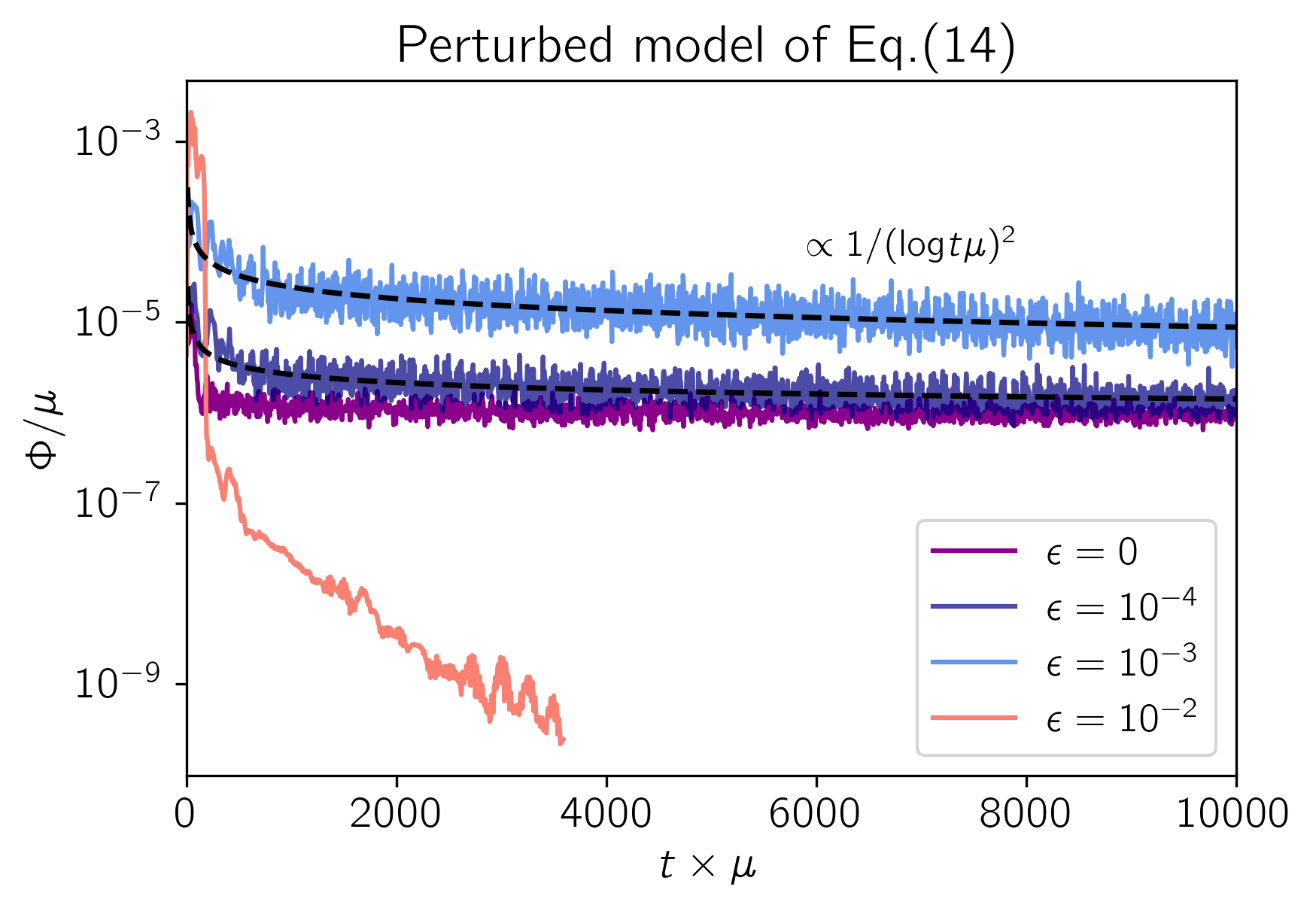}
    \caption{\textit{Left:} Behavior of $\Phi$ throughout the evolution of star \texttt{S005} with perturbation type~\eqref{eq:perturbation} varying both amplitude $\epsilon$ and azimuthal dependence characterized by $n$.
    The perturbation with $\epsilon = 0.1$ results in BH formation, signaled by a sudden drop of $\Phi$. \textit{Right:} Same as the left panel but with perturbation of type~\eqref{eq:perturbation_lr}; BH formation is now triggered for $\epsilon = 0.01$. Black dashed lines indicate the logarithmic-in-time decay of $\Phi$.}
    \label{fig:perturbation_example}
\end{figure*}

To assess the dynamical stability of ultracompact boson stars, we performed a series of evolutions of model \texttt{S005},
as well as other ultracompact stars that we summarize in Appendix~\ref{app:other_models}.
We find no sign of unstable (linear or nonlinear) behavior for solution \texttt{S005}, neither with, nor without additional perturbations (so long as their amplitude is sufficiently small to prevent immediate gravitational collapse) on the timescales quoted in Ref.~\cite{Cunha:2022gde}. Further, we isolate both numerical artifacts and linear gauge instabilities, which may lead to a misidentification of a physical linear or nonlinear departure from stationarity, and provide guidance to mitigate such issues. To that end, we begin by first presenting our results utilizing the setups shown in Table~\ref{tab:codes_setup}, and return to the numerical and gauge issues we encountered, as well as the main take-aways to address the stability of these extreme models, afterwards.

\subsection{Main stability results}

We start our discussion with three-dimensional evolutions of model \texttt{S005}, which we first perturb solely with the discretization error of our numerical methods. We have evolved\footnote{Although our lower resolution runs have been evolved to even longer timescales of $t\mu \sim 2 \times 10^{4}$.} this model up to $t\mu \sim 10^{4}$
using both codes, which support stability of these ultracompact spacetimes on the considered timescales. We further observe no signs of a departure away from the initial stationary BS solution. As an example, in Fig.~\ref{fig:stationarity}, we illustrate how this stationarity of the solution is preserved up to numerical errors and all $\Phi_{\tilde{m}}$ modes up to $\tilde{m} \leq 20$ remain roughly constant or decay throughout the evolution\footnote{At early times, our diagnostics oscillate more rapidly as a result of gauge dynamics.}. We note, however, that the $\tilde m = 4$ mode, normalized with respect to the background value of $|\Phi_0|$ displays a slow linear growth. This is likely an effect of anisotropies and irregularities introduced by the choice of our Cartesian AMR and regridding that happens at later times. Repeating the same simulation but with less irregular AMR grid reveals a suppression in the growth of the $\tilde m = 4$ mode.
Finally, we have also verified consistency of all $\tilde{m}$-modes across resolutions, finding convergence compatible with the expected fourth order. A discussion of the evolutions of \texttt{S005} using {\sc GHC} can be found in Appendix~\ref{app:gh_sims}; the behavior is qualitatively similar to the one shown here in the main text.
Note that in this case, with \textit{increasing} resolution, the amplitude of the initial perturbation \textit{decreases} (see,~e.g.,~Fig.~\ref{fig:gh_results}). 

By naive expectation, any nonlinear process driving the system away from the initially stationary solution (and its associated timescale) 
scales with the amplitude of the perturbation. As a result, perturbations deep in the linear regime (such as the ones present in Fig.~\ref{fig:stationarity}) 
may activate such nonlinear process only on much longer timescales. To circumvent this suppression and probe the stability of the \texttt{S005} model to large initial perturbations, we proceed by explicitly adding two types of such perturbations. Note that neither of the perturbations described below solve the Einstein constraint equations.\footnote{In fact, for large-amplitude perturbations we find a reduction of the convergence order of constraint violations, as measured by Eq.~\eqref{eq:ham_mom_constraints}.}

First, we investigate whether the model would suffer from any non-axisymmetric instabilities by introducing an azimuthal dependence to the initial perturbation and modifying the scalar field by
\begin{equation} \label{eq:perturbation}
    \varphi \rightarrow \varphi e^{-2\epsilon\mathrm{cos}(n \phi)}.
\end{equation}
Above, $\epsilon$ denotes the magnitude of the perturbation and $n$ an azimuthal number. Choosing $n$ in Eq.~\eqref{eq:perturbation} sets the leading contribution\footnote{Working under the assumption that $\epsilon \ll 1$, the expansion of the exponential term would result in an infinite sum of powers of $\mathrm{cos}(n \phi)$ terms for $n \geq 1$.} to the first few $n$-modes in the expansion in $\epsilon$. Consequently, we expect the excited $n$-modes to show up in the azimuthal decomposition of $\Phi$ in Eq.~\eqref{eq:decomposition}. Explicitly, we consider the evolution of star \texttt{S005} under this non-axisymmetric perturbation with $n = 1$ and $n = 2$ and varying amplitude $\epsilon$. In the left panel of Fig.~\ref{fig:perturbation_example}, we monitor the stationarity of the solution using $\Phi$ throughout the evolution of these initial data. With perturbation amplitudes of $\epsilon \geq 0.01$, the expected modes get excited more visibly and acquire an order of magnitude larger amplitudes than their unperturbed counterparts (see,~e.g.,~the left panel of Figure~\ref{fig:modes_perturbed_models} of Appendix ~\ref{app:additional_modes}, where we present the corresponding mode decomposition). In particular, for perturbed initial data with $n=1$ ($n=2$), the dominant modes in $\Phi$ we excite are $\tilde{m}=1$ and $\tilde{m}=2$ ($\tilde{m}=2$ and $\tilde{m}=4$). Physically, in the former case, the perturbations induce a kick on the star and cause bar-like deformations at later times\footnote{The star acquires the shape of a rugby ball.}, whilst in the latter, the bar-like deformations start to rotate in the equatorial plane. BH formation is only triggered for extortionately large values of $\epsilon \geq 0.1$ (see the left panel of Fig.~\ref{fig:perturbation_example}).

Secondly, we introduce a Gaussian perturbation directly around the position of the star's stable LR (with radial coordinate location $r_0$) 
\begin{equation} \label{eq:perturbation_lr}
    |\varphi| \rightarrow |\varphi| + \epsilon e^{(r-r_0)^2/\Delta^2}.
\end{equation}
Here $\Delta$ denotes the width of the Gaussian profile and, for simplicity, we fix $\mu \Delta = 1$.
In this case the perturbation generally results in an excited star that starts to oscillate radially around its equilibrium state. These induced oscillations are long-lived, as can particularly be seen from $\Phi$ in the the case of larger amplitude $\epsilon = 10^{-3}$ (see the right panel of Fig.~\ref{fig:perturbation_example}).
Analogous to the previous perturbation, BH formation occurs only for sufficiently large amplitude ($\epsilon \geq 0.01$ in this case),
and the amplitude of the initial perturbation of \texttt{S005}, as measured by $\Phi$, is orders of magnitude larger between the cases with explicit perturbation~\eqref{eq:perturbation_lr} and the one without, as can be seen by comparing Figs.~\ref{fig:perturbation_example} and~\ref{fig:stationarity}. 

All in all, despite exciting solution \texttt{S005} with as large a perturbation as possible (before inducing prompt collapse), we are unable to identify a late-time drastic non-perturbative departure from the initial (perturbed) stationary solution. 
It is, however, possible that an unstable nonlinear process is triggered on the timescale considered here only for perturbations with amplitudes between the ones that lead to prompt collapse and those inducing long-lived, but decaying, perturbations, or by a different type of perturbation altogether. 

\begin{figure}[t!]
    \centering
    \includegraphics[width=1\linewidth]{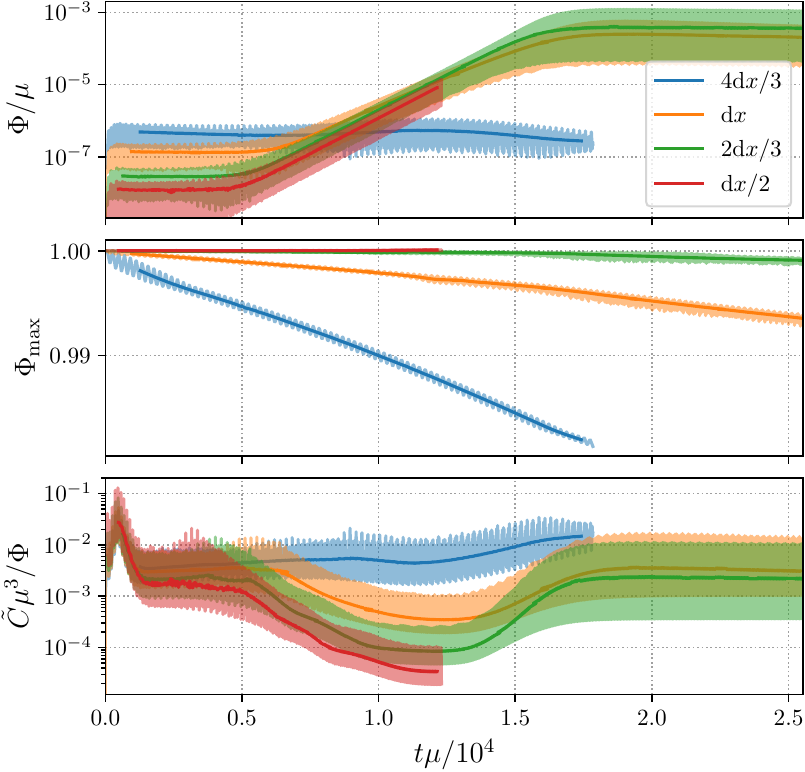}
    \caption{Evolution of the ultracompact spinning boson star \texttt{S02} that exhibits a gauge instability in stationary gauge using the {\sc GHC}. The solid lines are moving averages of each diagnostic (shaded lines). \textit{Top panel:} Maximum of the derivative of the scalar field amplitude, as defined in Eq.~\eqref{eq:phi_diagnostics}; $\mu \du x$ is the grid spacing of the reference resolution quoted in the main text. \textit{Middle panel:} The ratio of the maximum of the scalar field amplitude to its value at initial time $t=0$, as defined in Eq.~\eqref{eq:phi_max}. \textit{Bottom panel:} Integrated constraint violations, defined in Eq.~\ref{eq:gh_cnst}, and normalized by $\Phi$.
    }
    \label{fig:gauge_instability}
\end{figure}

\subsection{Unphysical instabilities}

\subsubsection{Gauge instabilities}\label{sec:gauge_generalised}

Beyond potential nonlinear instability mechanisms, fully nonlinear evolutions can be (and have been) used to identify linear dynamical instabilities. These manifest as exponentially growing (linear) perturbations away from a background spacetime. In contrast to any nonlinear process, this unstable behavior is amplitude-independent (e.g., an amplitude-independent instability timescale can be defined). To assess the presence of such an instability using nonlinear evolutions, one would naively expect that diagnostics, such as Eqs.\eqref{eq:phi_diagnostics}--\eqref{eq:gtt_diagnostics}, and \eqref{eq:decomposition}, exhibit exponential growth. In the context of General Relativity, such growth would need to be not only consistent, but also exhibit converging Einstein constraint violations, across different resolutions\footnote{Numerical instabilities typically grow exponentially too. However, their growth rate is resolution-dependent and violates the constraints.}. 

While the gauge-dependent diagnostics in Eqs.\eqref{eq:phi_diagnostics}--\eqref{eq:gtt_diagnostics},\eqref{eq:decomposition} can be useful guides for assessing instability and quantifying their nature, they \textit{may not} necessarily imply a \textit{physical} instability. To illustrate this, we present the nonlinear evolution of star \texttt{S02} in Fig.~\ref{fig:gauge_instability} in axisymmetry; that is, we impose axisymmetry on the metric and $\partial_\phi\varphi=3i\varphi$, and use a reference resolution of $\mu \du x \approx 0.003$ on the finest level. To that end, we employ the generalized ``cartoon'' method (see e.g.~\cite{Pretorius:2004jg,Alcubierre:1999ab,Cook:2016soy}) within the {\sc GHC} framework, complemented by the stationary gauge. In this setup, the solution \texttt{S02} is expected to be \textit{linearly} stable.\footnote{The axisymmetric setting discussed here suppresses the physical non-axisymmetric instability found in Refs.~\cite{Sanchis-Gual:2019ljs,Siemonsen:2020hcg}.} Consulting Fig.~\ref{fig:gauge_instability}, we find that both $\Phi$ and $C$ converge to zero at early times (at roughly the expected fourth order). However, after $\mu t \sim 5000$, 
we find a gauge instability to emerge and drive an exponential growth of $\Phi$ (as well as $G$, although not shown here). Importantly, during this unstable phase the constraint violations converge to zero at the expected order, and the e-folding time of $\Phi$ converges to a \emph{non-zero} value with resolution (see the upper panel of Fig.~\ref{fig:gauge_instability}). Throughout the exponential growth and saturation of the instability, the \emph{gauge-invariant} quantity $\Phi_{\max}$ remains constant, up to a linear drift and small oscillations, which converge away. The endstate of the gauge instability we found to be radial oscillations of $\varphi$ with respect to the coordinate domain. This illustrates that identifying a physical linear instability requires not only converging exponential growth in gauge-dependent quantities such as $\Phi$ and $G$, but further gauge-invariant validation. 
For instance, in Ref.~\cite{Siemonsen:2020hcg} a linear instability was identified using $\Phi$ and $G$, but further verified with the behavior of $\Phi_{\rm{max}}$, 
particularly once the unstable modes entered the nonlinear regime. In Appendix~\ref{app:gauge_instability}, we provide further details on gauge instabilities in scalar BSs within the generalized harmonic formulation; in particular, we explicitly demonstrate that the instability timescale scales with the compactness of the background star solution. Lastly, due to the symmetry assumptions, and the absence of polar LRs\footnote{We define the polar LR to be the collection of trapped zero-angular momentum null geodesics.} 
in \texttt{S02}, a nonlinear instability associated with trapped gravitational modes in the LRs is not expected to be present here.

\begin{figure}[t!]
    \centering
    \includegraphics[width=1.0\linewidth]{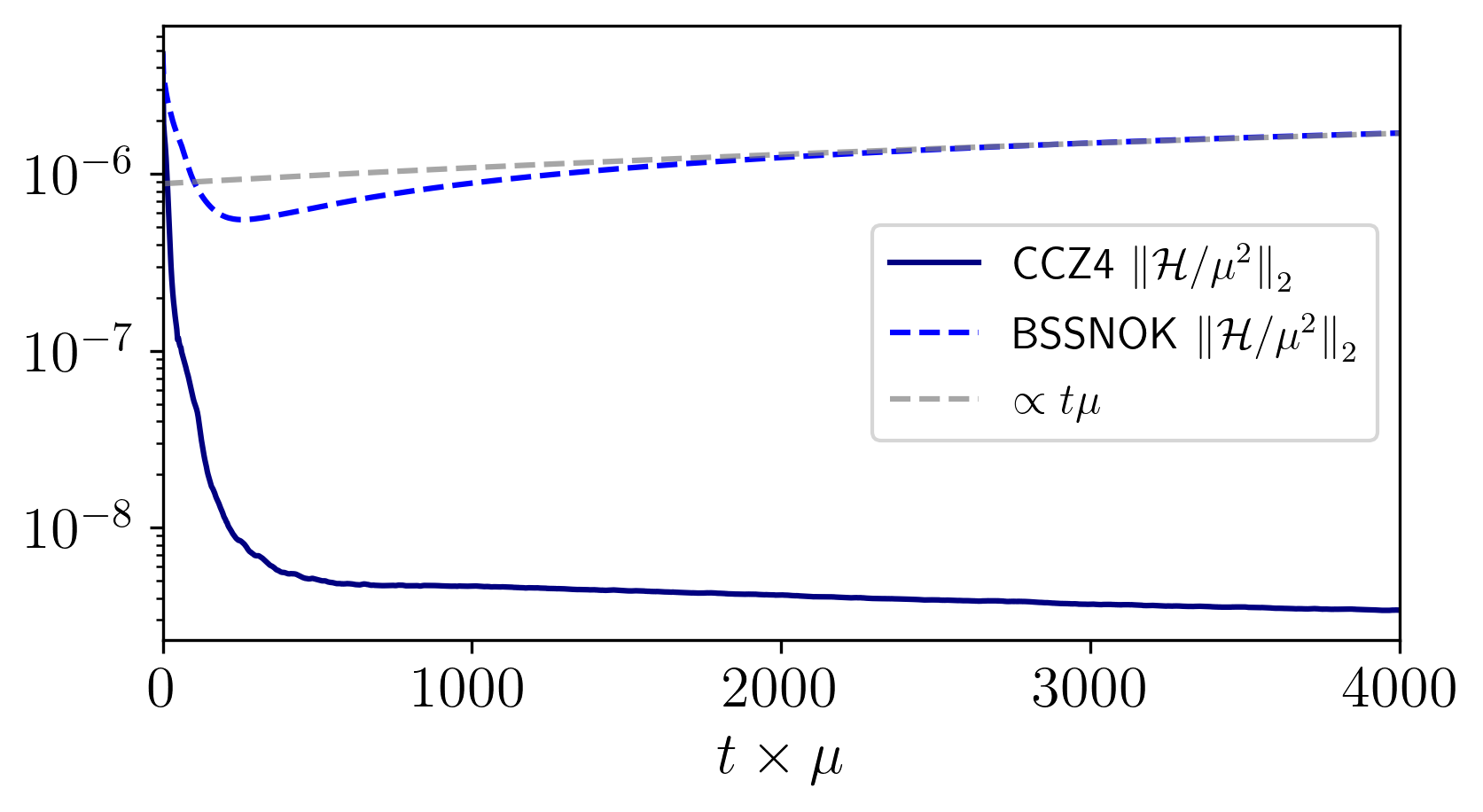}
    \caption{A zoom-in on volume-weighted $L^2$-norms of Hamiltonian constraint for the \texttt{S005} BS model evolved with $\mu \du x \approx 0.083$. We compare the evolution of constraints using CCZ4 and BSSNOK formulations, where the latter evidently exhibits problematic growth. 
    }
    \label{fig:bssn_and_ccz4}
\end{figure}

\subsubsection{Constraint damping}

As first pointed out in Ref.~\cite{Marks:2025jpt}, we have also found that constraint damping plays a key role in long-term evolutions using the {\sc ExoZvezda} code. In particular, in addition to the results presented above obtained using the CCZ4 formulation, we evolved the \texttt{S005} model in the exact same set-up of Table~\ref{tab:codes_setup} (and identical gauge), but with no constraint damping using the BSSNOK formulation~\cite{Baumgarte:1998te,Shibata:1995we,Nakamura:1987zz}. All of our BSSNOK simulations, however, suffered from a gradual growth in constraint violations, as illustrated in~Fig.~\ref{fig:bssn_and_ccz4}, reminiscent of constraint growth reported in neutron stars evolutions in Ref.~\cite{Bernuzzi:2009ex} (see their Fig.~5). The BSSNOK constraint subsystem has previously been shown to possess a zero-speed characteristic (see e.g.~Section VI of Ref.~\cite{Beyer:2004sv} or Section II of Ref.~\cite{Brown:2008sb}), which can accumulate in time and lead to large constraint violations during the evolution. Even at the initial time we inevitably accrue some constraint violations due to interpolation errors and imperfections in initial data\footnote{We note that we achieve first order point-wise convergence of the Hamiltonian constraint at the initial time slice.}; particularly around the regions of the star where the scalar field gradients are large. In BSSNOK simulations, however, these constraint violations remain stationary on the grid and gradually start to accumulate in time, whilst in CCZ4 such constraint violating modes propagate away quickly. In fact, in BSSNOK runs we observe a deterioration in the convergence of the $L^2$-norm of the constraints to only first order during the evolution.
Furthermore, CCZ4 reduces the Hamiltonian constraint violations by two orders of magnitude and the drift in $\Phi_{\rm{max}}$ by one order of magnitude.  

\section{Conclusions}\label{sec:conclusions}

In this work, we revisited the question of the stability of ultracompact spinning BSs with a stable LR, investigated in Ref.~\cite{Cunha:2022gde}, using 3+1 numerical relativity simulations. 
We perturbed these stationary solutions by both discretization error of our numerical methods and two different types of explicit perturbations of varying initial amplitude and azimuthal dependence, and further used both the CCZ4 and generalized harmonic formulations of the Einstein equations. We found no evidence of conjectured light ring instabilities on timescales of $t\mu \sim \mathcal{O}(10^4)$, whereby we monitored several diagnostics and ensured that the spacetime at hand remained close to its initial stationary state. We did find spinning ultracompact spacetimes to be significantly more challenging to treat using numerical relativity; more care was needed to cure potential non-physical instabilities, stemming from gauge or evolution scheme choices. We therefore also summarized the key ingredients that facilitated our long-term successful evolutions. 

We stress that our study does \textit{not} provide a self-contained proof of the non-linear stability of these types of BS spacetimes. On the contrary, more investigations are needed to fully understand the impact of slow logarithmic decay observed in e.g. Ref.~\cite{Keir:2014oka} at the linear level on the fully non-linear problem. Given the relatively large stability timescale of our numerical simulations, the character of the instability suggested in Ref.~\cite{Keir:2014oka}, if present at all, is likely to be very weak. In particular, if massless modes were to get trapped inside the light ring region, their accumulation may not manifest itself for a long time. 
Even if the nonlinear interactions do not ultimately lead to instability, it is interesting to investigate the impact of stable null trapping on the spacetime of ultracompact objects.
A way to address this question in future work may be to start explicitly with some perturbing matter or field configuration that at the test-field level would be trapped in the light ring (or exhibit slow decay), and then study its gravitational backreaction. 


\begin{acknowledgments}
T.E. thanks Ulrich Sperhake, Seppe Staelens, and Gareth Marks for useful discussions. Research at Perimeter Institute is supported in part by the Government of Canada through the Department of Innovation, Science and Economic Development and by the Province of Ontario through the Ministry of Colleges and Universities.
We acknowledge support by the Cambridge Service for Data Driven Discovery
at the University of Cambridge and
Durham University through DiRAC Projects ACTP284 and ACTP and STFC capital Grants No.~ST/P002307/1 and
No.~ST/R002452/1, and STFC operations Grant No.~ST/R00689X/1.
W.E. acknowledges support from a
Natural Sciences and Engineering Research Council
of Canada Discovery Grant and an Ontario Ministry
of Colleges and Universities Early Researcher Award.
This research was enabled in part
by support provided by SciNet (www.scinethpc.ca)
and the Digital Research Alliance of Canada (alliancecan.ca). 
Computations were done on the CSD3 (Cambridge), Cosma7 and 8 (Durham) and Niagara (University of Toronto) clusters. 
\end{acknowledgments}


\bibliographystyle{apsrev4-1}
\bibliography{bibliography}

\appendix
\section{Convergence} 
\subsection{{\sc ExoZvezda} code}
\label{convergence_tests}
\begin{figure*}[t!]
    \centering
    \includegraphics[width=1.0\linewidth]{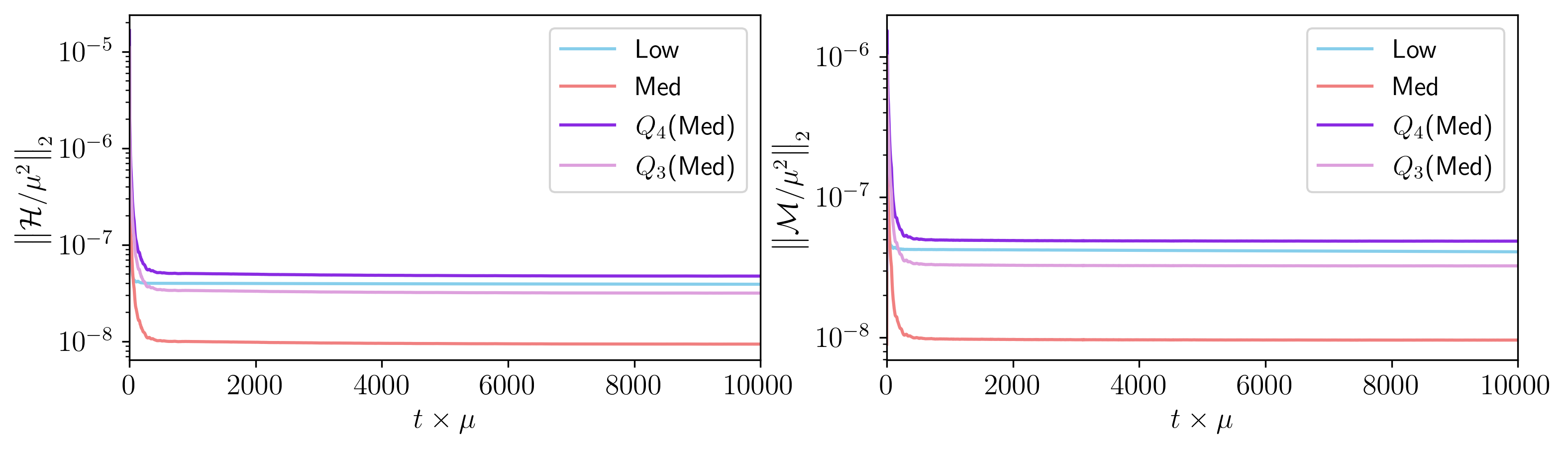}
    \caption{Convergence of the volume-weighted $L^2$-norms of the Hamiltonian (left) and momentum (right) constraints in {\sc ExoZvezda}. Medium (low) resolutions use $\mu \du x \approx 0.167 $ ($\mu \du x \approx 0.25 $) on the finest levels. We find consistent convergent behavior in the constraints, sandwiched between third and fourth orders.}
    \label{fig:constraints_conv}
\end{figure*}

In this section, we summarize the results of the convergence analysis for the \texttt{S005} model, perturbed by the discretization error, using the {\sc ExoZvezda} code. We utilize $\mu \du x \approx 0.25$, $\mu \du x \approx 0.167$, and $\mu \du x \approx 0.125$ on the finest level for our low, medium, and high resolutions, respectively. As illustrated in Fig.~\ref{fig:constraints_conv}, we find between 3rd and 4th order convergence for the Hamiltonian and momentum constraints. We have also evolved this model at even higher resolutions of $\mu \du x = 0.083$ and $\mu \du x = 0.042$, albeit on shorter timescales, and find the behavior of diagnostics consistent with the resolutions employed in our convergence test.

\subsection{Generalized harmonic simulations} \label{app:gh_sims} 
\begin{figure}
    \centering
    \includegraphics[width=1.0\linewidth]{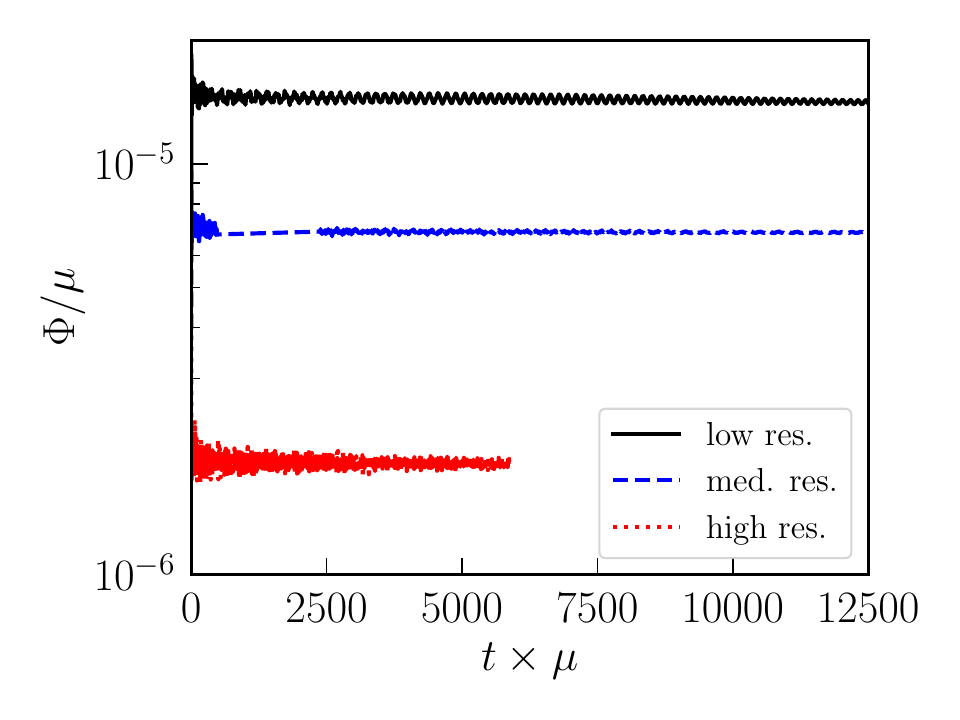}
    \caption{The maximum of the absolute value of $\partial_t |\varphi|^2$ over
    the domain (see~Eq.~\eqref{eq:phi_diagnostics}), as a function of time for evolutions of the rotating BS
    solution \texttt{S005} carried out in the generalized harmonic formulation.  The three
    different curves correspond to different numerical resolutions and
    consistent with converging to zero at approximately third order. We note that there is some missing data between $\mu t \sim 500-2000$ for the medium resolution run.}
    \label{fig:gh_results}
\end{figure}

We have also evolved the \texttt{S005} model in 3+1 using a generalized harmonic
formulation and the methods and code described in Ref.~\cite{Siemonsen:2020hcg} (see~also Section~\ref{sec:numerics}).
We find no evidence for an instability in the BS on timescales of $\sim
10^4 \mu^{-1}$. Here, we fix the gauge by requiring the source functions $H^a =
\Box x^a $ to be constant in time and equal to the those of the stationary
boson star solution, and let the solution be perturbed only by discretization error.
In Fig.~\ref{fig:gh_results}, we show $\Phi$ over
the domain, as a function of time for simulations with three numerical
resolutions. The lowest resolution has a grid spacing of $\du x \approx 0.3
\mu^{-1}$ on the finest level, while the medium and high resolutions have grid spacings that are,
respectively, $\times 3/4$ and $\times 1/2$ as large.  There is no noticeable
growth in this quantity over the timescales covered.  As expected for a
stationary solution perturbed only by discretization error, the overall value is
converging to zero with increasing resolution. In these runs we did not see the appearance of a gauge instability, but considering the same case and restricting to axisymmetry, we do see a similar gauge instability to the one described in the main text and Appendix~\ref{app:gauge_instability} appear on timescales a factor of a few longer.

\section{Additional mode decomposition results}\label{app:additional_modes}

\begin{figure*}[t!]
    \centering
    \includegraphics[width=0.45\linewidth]{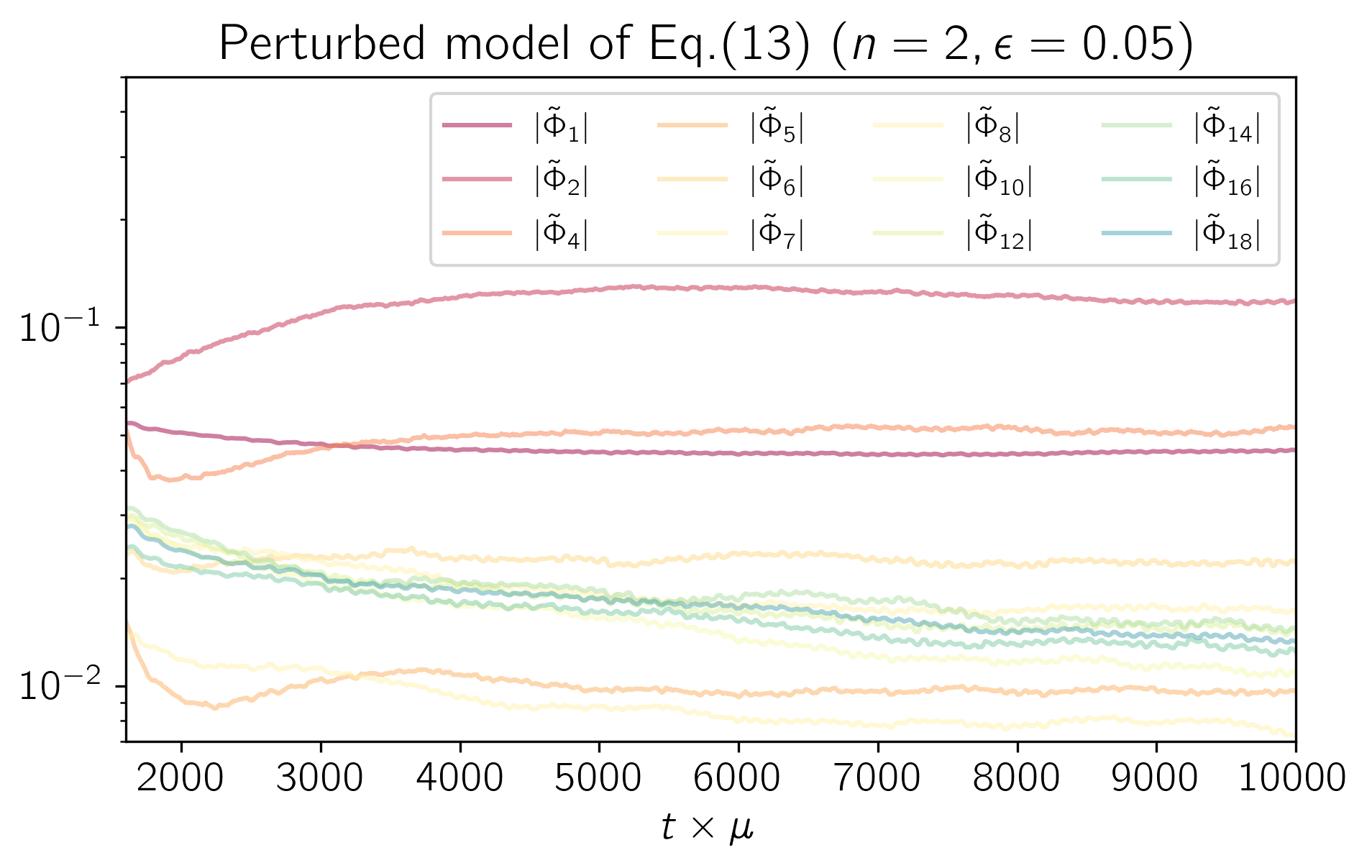}
    \includegraphics[width=0.45\linewidth]{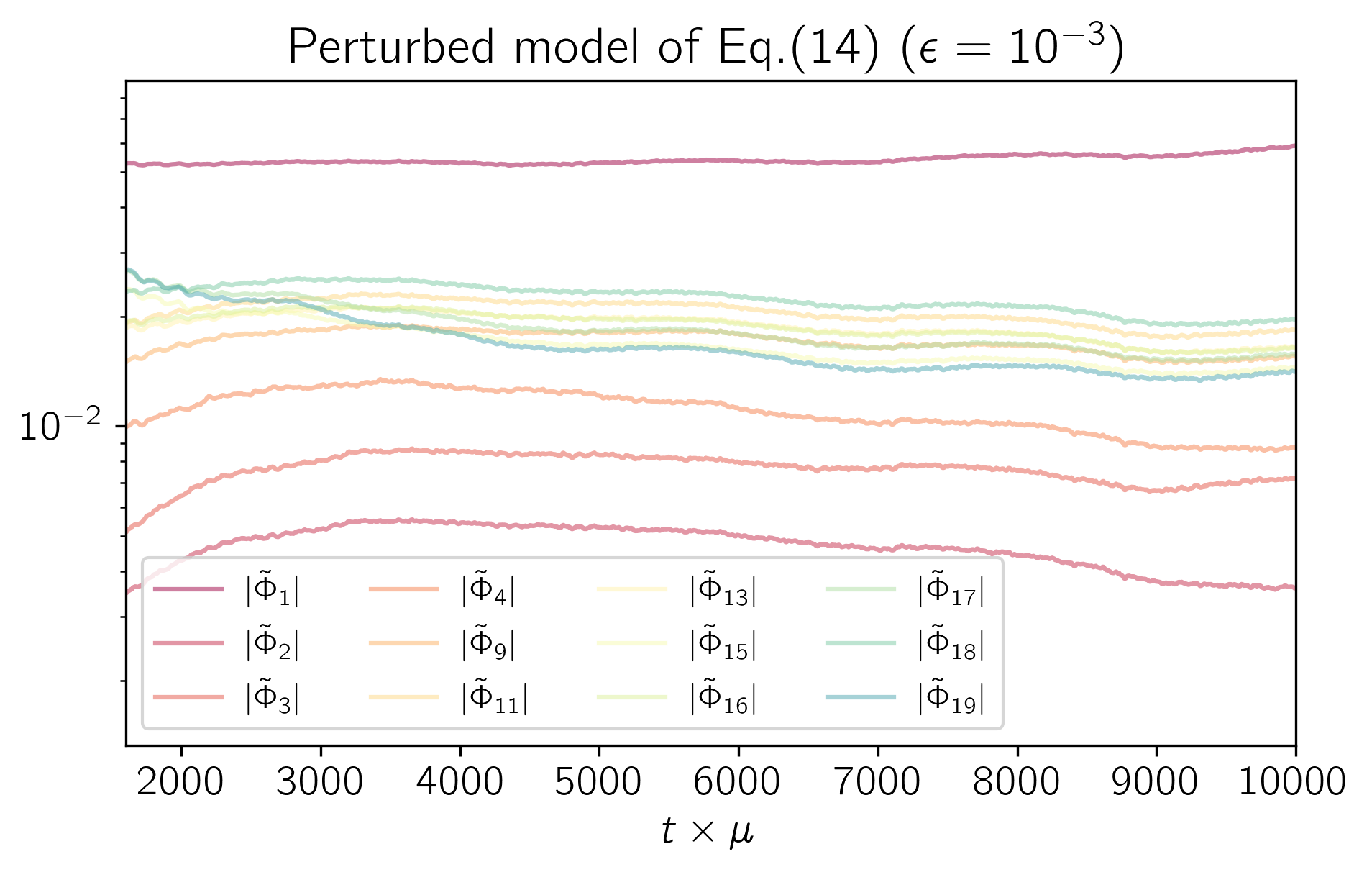}
    \caption{Representative examples of mode behavior of the \texttt{S005} model using perturbed initial data of Eq.~\eqref{eq:perturbation} (left panel) and \eqref{eq:perturbation_lr} (right panel) investigated in the main text. We remind the reader that we plot the normalized power in the modes, i.e.~$|\tilde{\Phi}_{\tilde{m}}| = |\Phi_{\tilde{m}}|/|\Phi_0|$. For visual purposes, we represent this quantity as a rolling time-average with a window of $\mu t \sim 1600$. 
    }
    \label{fig:modes_perturbed_models}
\end{figure*}

In the main text, we have presented the mode decomposition of $\Phi$ according to Eq.~\eqref{eq:decomposition} for the evolution of the \texttt{S005} model solely driven by discretization error (see Fig.~\ref{fig:stationarity}). In this section, we supplement our mode decomposition results for the \texttt{S005} model, but now subjected to initial perturbations given by Eqs.~\eqref{eq:perturbation} and \eqref{eq:perturbation_lr}. We remind the reader that the initial perturbation of Eq.~\eqref{eq:perturbation} incorporates azimuthal mode dependence via a choice of $n$, which is a priori arbitrary. In the left panel of Fig.~\ref{fig:modes_perturbed_models}, we illustrate the behavior of modes for this perturbation with $n=2$ and $\epsilon = 0.05$. As expected from this type of perturbation, the leading modes that are excited in $\Phi$ are the $\tilde{m} =2$ and $\tilde{m} = 4$ modes. On the other hand, Eq.~\eqref{eq:perturbation_lr} introduces a Gaussian-like type of perturbation around the light-ring region. As illustrated in the right panel of Fig.~\ref{fig:modes_perturbed_models}, we excite a variety of modes with some of the higher $\tilde{m}$-modes acquiring larger amplitudes. Overall, with both of the initial perturbations we consider, the excited modes do not exhibit any conspicuous growth.

\section{Gauge instability} \label{app:gauge_instability}

\begin{figure}[b]
    \centering
    \includegraphics[width=1\linewidth]{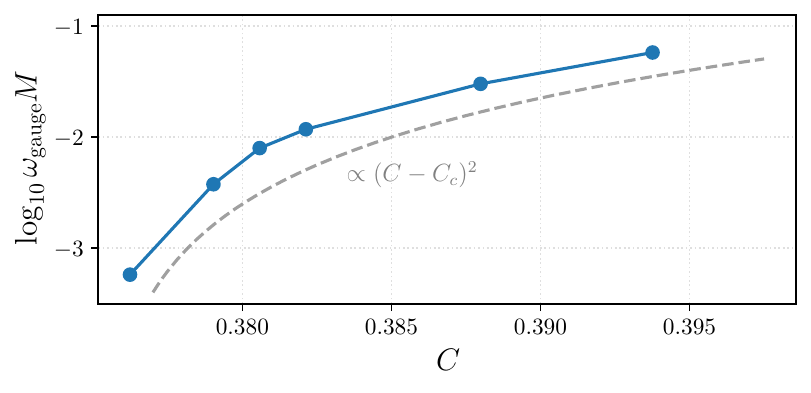}
    \caption{Growth rate $\omega_{\rm gauge}$ of the (most unstable) gauge instability 
    as a function of the compactness $C$ of solutions along the family of solitonic $m=3$ spinning BSs with $\sigma_0=0.2$. The gray dashed line indicates a $\sim C^2$ scaling that the rate follows above a some critical $C_c$. 
    }
    \label{fig:gauge_inst_rates}
\end{figure}

Here we provide further details on the gauge instability in scalar BSs within the generalized harmonic framework of the {\sc GHC} code (see Table~\ref{tab:codes_setup}). While our discussion in the main text was focused on the solution \texttt{S02}, we find that the gauge instability is present along the $m=3$ family of rotating BSs in the solitonic model with coupling $\sigma_0=0.2$. In Fig.~\ref{fig:gauge_inst_rates}, we present the growth rate $\omega_{\rm gauge}$ of this instability along this family of solution parameterized by their compactness $C=M/R$. This instability appears above a critical compactness and the growth rate begins increasing quadratically with $C$. Furthermore, we find this instability not only in spinning stars with solitonic self-interactions, but also in scalar models with axionic potentials and leading repulsive self-interactions (see Ref.~\cite{Siemonsen:2020hcg} for details on these potentials); a similar behavior was observed also in evolutions of highly compact Proca stars with ergoregions. Additionally, we have observed a very slowly growing gauge mode in highly compact, $C\approx 0.3$, spherically symmetric BSs in the solitonic scalar model with $\sigma_0=0.05$. Lastly, on long time scales, accessible only within this axisymmetric setup, we also find the \texttt{S005} model (discussed in detail in Appendix~\ref{app:gh_sims}) to suffer from an analogous gauge instability. 
From this, we conclude that the stationary gauge within the generalized harmonic formulation (and assuming the form of Eq.~\eqref{eq:metric}) is unstable for a large class of highly compact spacetimes (independently of the matter source and presence of an ergoregion). 

The nonlinear development of these gauge instabilities we classify into three qualitatively different behaviors: (i) saturation with rapid radial oscillations of the star characterized by a few frequencies, 
(ii) saturation through a drift of the solution away from the initial coordinate location, and (iii) likely encountering a coordinate singularity, beyond which, numerical integration fails.

We have succeeded in removing this instability \textit{completely} within the generalized harmonic framework only in the case of highly compact spherically symmetric BSs. In general, we find the CCZ4 formulation (and the associated puncture gauges) to be substantially more robust in stably evolving highly compact (spinning as well as nonspinning) BSs. For future reference, in the remainder of this appendix, we outline our approach to stabilize the gauge for evolutions of non-spinning BSs and briefly list our approaches to address (and partially mitigate) this instability for spinning BSs within the generalized harmonic formulation.

\subsection{Spherically symmetric boson stars}

We are able to completely remove the gauge instability we found to be present in the $m=0$ BS with frequency $\omega/\mu=0.122$ in the solitonic scalar model with $\sigma_0=0.05$. To that end, we transform from the metric of the form~\eqref{eq:metric} using
\begin{align}
x=k(r)\sin\theta\cos\phi, & & y=k(r)\sin\theta\sin\phi, & & z=k(r)\cos\theta,
\end{align}
to Cartesian coordinates. Here $k(r)$ is an invertible function over $r\in(0,\infty)$. We then impose the Cartesian harmonic gauge condition on the metric, in these new coordinates $x'^\mu=(t,x,y,z)^\mu$ we have
\begin{align}
\square x'^\mu=0.
\label{eq:harmoniccond}
\end{align}
In the condition~\eqref{eq:harmoniccond}, all spatial components result in the same ordinary differential equation for the unspecified function $k(r)$
\begin{align}
\partial_r^2 k(r)+\frac{2\partial_r k(r)}{r}+\frac{[\partial_r k(r)][\partial_r l(r)]}{2l(r)}-\frac{2k(r)}{r}=0,
\end{align}
where $l(r)$ enters \eqref{eq:metric} (note that for $m=0$, the metric functions are independent of $\theta$). For numerical convenience, we introduce the field redefinition $k(r)=r \Gamma(r)$ and impose the boundary conditions: $\Gamma(r=0)=1$ and $(\partial_r \Gamma)(r=0)=0$. We construct the function $k(r)$, and hence its inverse $r=k^{-1}(\sqrt{x^2+y^2+z^2})$, by numerically integrating the above differential equation from the origin outwards. With this, we can use the metric in the new coordinates to start a nonlinear numerical evolution of a non-spinning BS in Cartesian harmonic coordinates. Note, for rotating spacetimes, the transformation to harmonic coordinates generally requires solving a two-dimensional elliptic partial differential equation; therefore, we restrict to the spherically symmetric setting here. 
With this, a stable evolution is achieved by using $H^a=H^a_{\rm DH}(1-f)$,
where $f$ is a time dependent function that evolves from $f=1$ down to $0$, to dynamically impose the damped harmonic gauge~\cite{Choptuik:2009ww,Lindblom:2009tu}; here $H^a_{\rm DH}$ are the standard damped harmonic source functions. We also explored  dynamically transitioning from the stationary gauge (assuming Eq.~\eqref{eq:metric}) to damped harmonic gauge directly. However, in all considered cases (spinning and non-spinning), we were unsuccessful due to the violent gauge dynamics. 

\subsection{Axisymmetric boson stars}
The inclusion of a gauge damping term in the generalized harmonic source functions, of the form
\begin{equation}\label{eq:lapsedampened}
    n_\mu H^\mu=\mu_L(\alpha-\hat{\alpha})/{\alpha^2},
\end{equation}
where $\mu_L$ is a damping rate, $\alpha$ is the lapse, $\hat{\alpha}\defeq \alpha(t=0)$, and $n_\mu$ is the unit normal to the spatial hypersurface, only partially mitigated the gauge instability observed in spinning boson stars. In particular, varying $\mu_L$, we were able to decrease the gauge instability's growth rate $\omega_{\rm gauge}$ by a factor of a few. In contrast, we also found that adding an analogous term to the shift's implied evolution equation increased the rate, i.e., worsened the instability. 

Lastly, the \textit{modified} generalized harmonic formulation \cite{Kovacs:2020ywu} (introduced as a well-posed formulation of certain modified gravity theories) makes available other gauge choices not present in the ordinary generalized harmonic framework. Following the notation of Ref.~\cite{Kovacs:2020ywu}, the cone-function $a(x)$ can be changed to modify the gauge evolution (fully within general relativity). Using the methods of Ref.~\cite{East:2020hgw}, for the most compact solutions we considered, we found the choice $a(x)=\alpha/2-1$ (together with stationary gauge for $H_\mu$ or lapse-damped stationary gauge, as in Eq.~\ref{eq:lapsedampened}) to give the best results in terms of addressing the instability, reducing the growth rate $\omega_{\rm gauge}$ by a factor of few or more for the \texttt{S02} model. However, even here, we were unable to completely remove the exponential growth of gauge modes.

\section{List of evolved models} \label{app:other_models}

\begin{figure}
    \centering
    \includegraphics[width=0.8\linewidth]{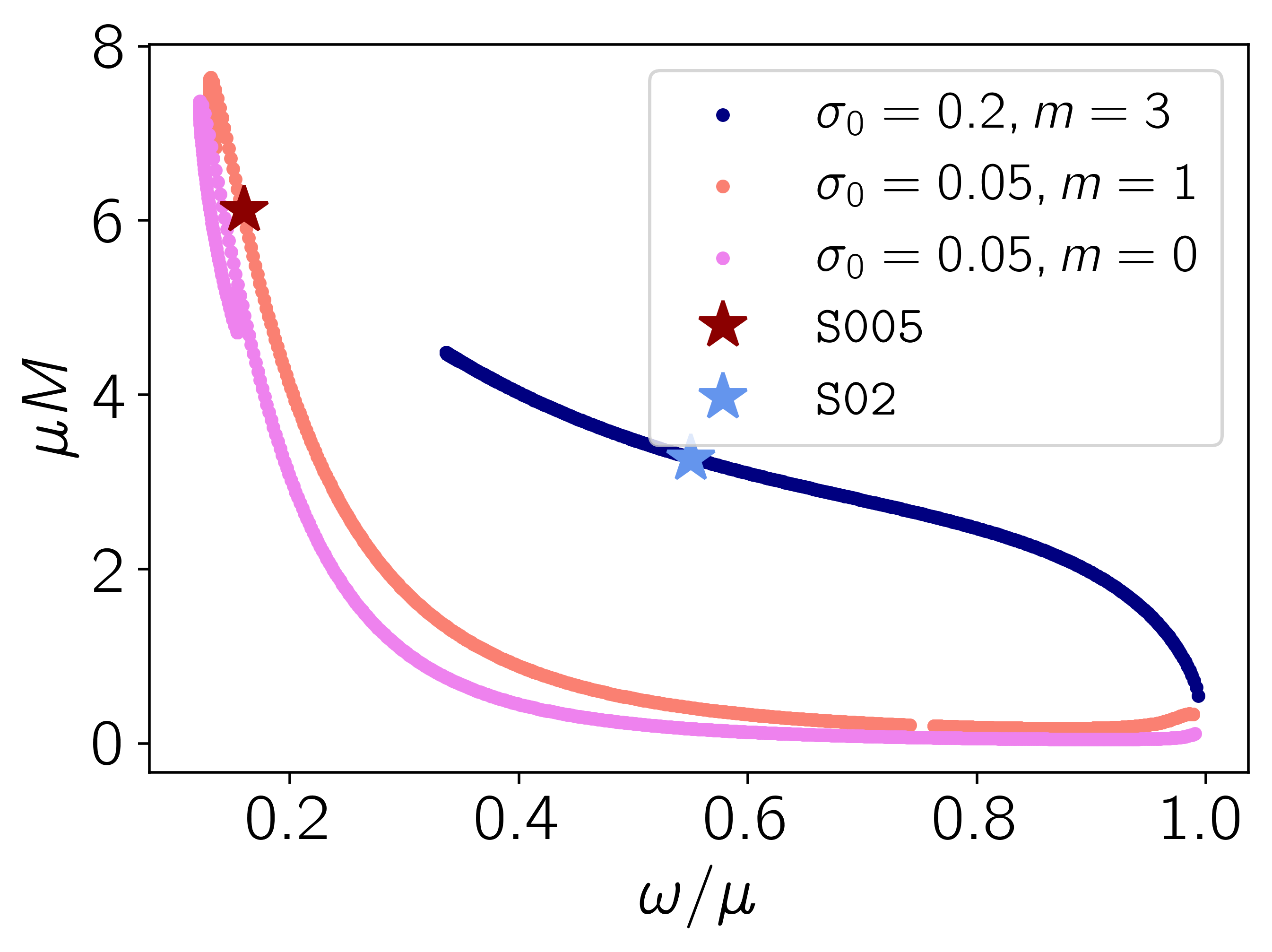}
    \caption{Main properties (ADM mass and BS frequency) of $\sigma_0 = 0.05, \, m=1$; $\sigma_0 = 0.05, \, m=0$; and $\sigma_0 = 0.2, \, m=3$ BS families featured in the Table~\ref{tab:allmodels}. We use the star markers to indicate the \texttt{S005} and \texttt{S02} solutions discussed in the main text.}
    \label{fig:fam-plot}
\end{figure}

In Table~\ref{tab:allmodels}, we summarize all the models (including the \texttt{S005} and \texttt{S02} featured in the main text) that we have investigated for stability. 
Further, in Fig.~\ref{fig:fam-plot} we display where in the parameter space these models reside.

\begin{table}[b!]
\centering
\caption{Summary of all models we have evolved in this work. We list the code name, number of dimensions of the evolution not constrained by symmetry, the azimuthal number of the scalar field, solitonic coupling, BS frequency, the ADM mass, compactness and the stability time-scale to which the model has been evolved to.
}
\begin{tabular}{| c | c | c | c | c | c | c | c |} 
\hline
~~Code~&~$d$~&~$m$~&~$\sigma_0$~&~$\omega/\mu$~&~$\mu M$~&~$C$~&~$\mu t$~\\ 
\hline
{\sc ExoZvezda} & 3 & 1 & 0.05 & 0.16 & 6.12 & 0.33 & $10^4$ \\ 
{\sc ExoZvezda} & 3 & 1 & 0.05 & 0.30 & 1.74 & 0.17 & $10^4$ \\ 
{\sc GHC} & 3 & 1 & 0.05 & 0.16 & 6.12 & 0.33 & $1.25\times10^4$ \\ 
{\sc GHC} & 2 & 1 & 0.05 & 0.16 & 6.12 & 0.33 & $3\times10^4$ \\
\hline
{\sc GHC} & 2 & 3 & 0.2 & 0.55 & 3.27 & 0.38 & $2.5\times 10^4$ \\
\hline
{\sc GHC} & 2 & 0 & 0.05 & 0.12 & 7.37 & 0.32 & $4\times 10^4$ \\
\hline
\end{tabular}
\label{tab:allmodels} 
\end{table}

\end{document}